\documentclass{aa}
\usepackage{graphics}
\usepackage{natbib}

\newcommand{\kms}{\hbox{km\,s$^{-1}$}}

\begin{document}

\msnr{accepted version}

\title{A spectral and spatial analysis of $\eta$ Carinae's 
diffuse X-ray emission using CHANDRA}

\author{Kerstin Weis \inst{1,2}\thanks{Visiting Astronomer, Cerro
Tololo Inter-American Observatory, National Optical Astronomy
Observatories, operated by the Association of Universities for
Research in Astronomy, Inc., under contract with the National
Science Foundation.}\fnmsep\thanks{Feodor-Lynen
fellow of the Alexander-von-Humboldt foundation}
\and Michael F.\ Corcoran \inst{3,4} 
\and Dominik J. Bomans\inst{1}
\and Kris Davidson \inst{2}}

\offprints{K.\ Weis, 
\email{kweis@astro.rub.de}}

\mail{K.\ Weis, Bochum, Germany}

\institute{Astronomisches Institut,
Ruhr-Universit\"at Bochum, Universit\"atsstr. 150, 44780 Bochum,
Germany
\and 
Astronomy Department, University of Minnesota, 116 Church Street SE,
Minneapolis, MN 55455, USA 
\and 
Laboratory for High Energy Astrophysics, Goddard Space Flight Center,
  Greenbelt, MD 20771, USA 
 \and 
 Universities Space Research Association, 7501 Forbes Blvd., Ste 206, 
Seabrook, MD 20706, USA}

\date{Received / Accepted}

\authorrunning{K.\ Weis et al.}
\titlerunning{$\eta$ Carinae's diffuse X-ray emission}

\abstract{The  luminous unstable star (star system) $\eta$ Carinae
is surrounded by an optically bright bipolar nebula, the {\it Homunculus} 
and a fainter but much larger nebula, the so-called {\it outer ejecta}. 
As images from the EINSTEIN and ROSAT satellites have shown, the 
outer ejecta is also visible in soft X-rays, while the central source is
present in the harder X-ray bands. With our CHANDRA observations we show that 
the morphology and properties of the X-ray nebula
are the result of shocks from fast clumps in the outer ejecta 
moving into a pre-existing denser circumstellar medium.
An additional contribution to the 
soft X-ray flux results from mutual interactions of clumps within the ejecta. 
Spectra extracted from the CHANDRA data yield gas 
temperatures $kT$ of $0.6-0.76$ keV. The implied 
pre-shock velocities  of 670-760\,\kms\ are within the scatter of the 
velocities we measure for the majority of the clumps in the corresponding 
regions.
Significant nitrogen enhancements over solar abundances are needed for 
acceptable fits in all parts of the outer ejecta, consistent with 
CNO processed material and non-uniform enhancement.
The presence of a diffuse spot of hard X-ray emission at the S\,condensation 
shows some contribution of the highest velocity clumps and further underlines 
the multicomponent, non-equilibrium nature of the X-ray nebula.
The detection of an X-ray ``bridge'' between the northern and southern part of
the X-ray nebula and an X-ray shadow at the position of the NN\,bow 
can be attributed to a large expanding disk, which would appear as an 
extension of the equatorial disk.  No soft emission is seen from the 
Homunculus, or from the NN\,bow or the ``strings''.
}
\maketitle

\section{Introduction}

\subsection{The object: $\eta$ Carinae\label{sec:eta}}

With luminosity $L\sim 10^{6.7}\,L_{\sun}$, $\eta$ Car marks 
the upper boundary of the empirical {\it Hertsprung-Russell diagram\/} (HRD). 
It is one of the most massive stars known 
($M > 120\,M_{\sun}$), somewhat evolved, and unstable;
see e.g.\ Humphreys \& Davidson (1994), Davidson \& Humphreys (1997) and 
Hillier et al. (2001).   

In the optical, $\eta$ Carinae is known to be an irregularly variable source. \
The historical lightcurve covers several centuries and shows the star's most
sudden change around  1843 when it brightened
to $-$1$^{\rm m}$ (Herschel 1847; Innes 1903; Viotti 1995; Humphreys et al.\ 
1999) and 
drastically decreased its brightness by more than  7$^{\rm m}$
during the following 20 years, an event known as the ``Great Eruption''.

$\eta$ Carinae is an evolved massive star, now classified 
a {\it Luminous Blue Variable\/} (LBV).
LBVs are very massive stars in a transitional phase between the 
main-sequence and the Wolf-Rayet state. Only stars 
above $\sim$  50\,M$_{\sun}$ enter this unstable phase, which occurs at an age of roughly 3 10$^6$ years (e.g.\ Langer et al.\ 1994). 
After the main-sequence phase such stars first 
evolve towards the red, cooler regime in the HRD, but as  they enter the LBV phase they
reverse their evolution before becoming red supergiants.
The temperature-dependent luminosity boundary in the HRD near which  LBVs are situated is known as the  
{\it Humphreys-Davidson Limit\/} (HD-Limit; Humphreys \& Davidson 1979, 1994).

A  periodicity first found in high excitation 
lines (Damineli 1996, Damineli et al.\ 1997) and in $2-10$ keV X-ray emission 
(Corcoran et al.\ 1995, 2001a; Ishibashi et al.\ 1999)
from the star raised the possibility that  $\eta$ Carinae is 
a binary system. In such a system a wind eclipse of a shock produced by the 
the colliding winds from two hot stars would be responsible for the observed 
X-ray variability. In this scenario, 
$\eta$ Carinae, the more massive component, has a mass 
in excess of  60\,M$_{\sun}$ while the companion would have a mass of about 
$30-60$\,M$_{\sun}$  (Damineli et al.\ 1997, Corcoran et al. 2001a). 
For further information concerning the high-energy X-ray emission
in connection with the central source we refer the reader to, for e.g.,\ 
Ishibashi et al.\ (1999), Corcoran et al.\ (2001a,b),  Pittard \& Corcoran (2002) and references therein. 

\subsection{The Homunculus nebula: optical and X-ray emission}

One of the main characteristics of the LBV phase is the star's 
optical variability,  which occurs on different timescales and amplitudes: from
a few tenths of a magnitude within several months, to $1-2$ magnitude variations on timescales of decades, to variations of several magnitudes as seen in the giant eruptions which occur 
probably once or twice during the LBV phase. Whether all LBVs undergo
giant eruptions is not clear. Therefore LBVs are sometimes divided
into the ``S Dor'' variables and the ``eruptive'' variables depending on whether or not an
eruption has taken place (Humphreys 1999).
$\eta$ Carinae is the canonical eruptive-type
LBV. Together with the high increase in mass loss which occurs 
during the LBV phase (several\,$10^{-4}$\,M$_{\sun}\,{\rm yr}^{-1}$ at least)  
these eruptions -- in which the stars peel of large parts of their outer 
envelopes -- are responsible for the formation of {\it LBV nebulae\/} (Nota et al.\ 1995; Weis 2001a).

\begin{figure}
{\resizebox{\hsize}{!}{\includegraphics{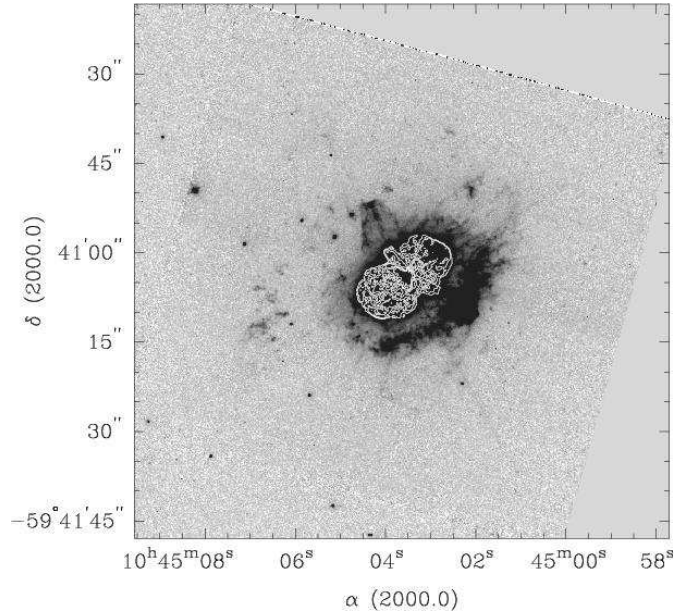}}} 
\caption{An HST F658N image which shows the optical emission from $\eta$
  Carinae's Homunculus and the outer ejecta. 
  The contrast has been optimized to show
  the fainter outer emission.  The bright central region including the 
  bipolar Homunculus is displayed using  contours.
}
\label{fig:hst}
\end{figure}

The bright ejecta-nebula around $\eta$ Carinae was first clearly 
identified in the 1940s 
(Gaviola 1946, 1950; Thackeray 1949, 1950) when it was about 30\%
smaller than today.  Gaviola called it the {\it Homunculus} because
it resembled a little man in his photographs.  At high spatial resolution, it 
has become one of the most dramatic 
examples of bipolar morphology, having two roughly symmetric polar 
lobes and an {\it equatorial disk} of material (Duschl et al.\
1995) which is sometimes called the skirt (e.g. Morse et al.\ 1998).  Its 
polar diameter is currently about 17\arcsec\ or 0.2\,pc.
Fainter condensations outside the Homunculus were 
also detected more than 50 years ago (Thackeray 1949; Walborn 1976),  
and manifest a much larger outer nebula, seen in
deep HST images (e.g.\ Weis 2001a) and images  in the soft X-ray energy  band. 
This {\it outer ejecta}, which consists of a large collection of 
``bullets'', ``knots'' and ``filaments'', has a diameter of about 60\arcsec\ or 
0.67\,pc. 
Figure \ref{fig:hst} displays an 
HST image taken in July 1997 in the F658N filter, which shows a 
mixture of H$_{\alpha}$ and [N\,{\sc ii}] emission.   Here the intensity 
levels are chosen to show the faint outer material, and we have 
indicated the location and size of the bipolar Homunculus
by contours.

\begin{figure}
{\resizebox{\hsize}{!}{\includegraphics{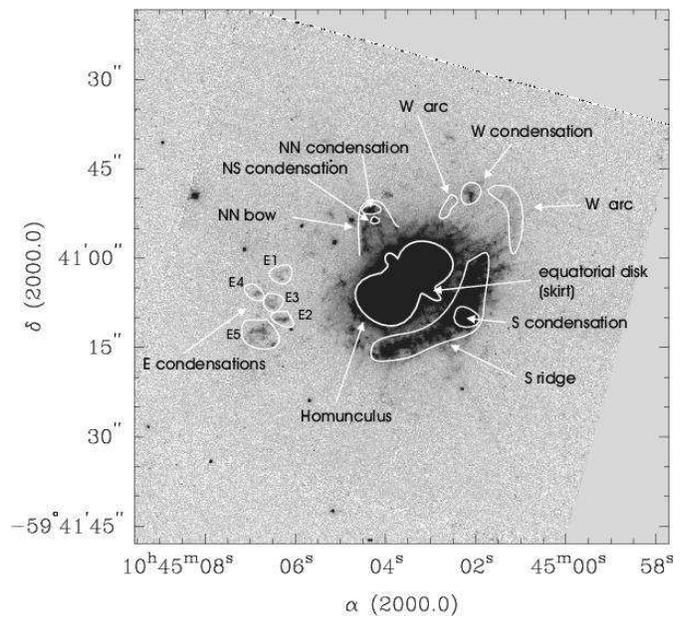}}} 
\caption{Previously identified features in the Homunculus and the outer ejecta are marked. 
Part of the nomenclature is according to Walborn et al.\ (1978)
}
\label{fig:name}
\end{figure}

Both parts of the nebula -- the Homunculus and the outer ejecta --
are expanding. The expansion velocity of the  Homunculus is about 650\,\kms\ 
(Thackeray 1961; Davidson \& Humphreys 1997) while in the outer ejecta
observed velocities average at roughly 600\,\kms\
-- slow enough to suggest either an earlier
ejection date or else deceleration  -- and reach maximum expansion 
velocities up to 2000\,\kms\ or more
(Weis 2001a,b; Weis et al.\ 2001). We will show in an upcoming paper that it
is indeed possible that the outer ejecta was  created 1843 and has been slowed down (Weis \& Duschl, in prep.).

The outer ejecta produces soft X-rays.
The EINSTEIN satellite separated the nebular from 
the stellar X-ray emission (Chlebowski et al.\ 1984), while
images with spatial resolution $\sim$ 5\arcsec\ were obtained later
with the High Resolution Imager (HRI) aboard the {\it R\"ontgensatellite},
ROSAT. Weis et al.\ (2001) presented a detailed analysis of the ROSAT data
and a comparison with optical images and kinematic data.  The X-ray
nebula is hook-shaped, with X-ray bright
knots near the locations of the prominent
{\it S\,ridge} and  {\it W\,arc} (features named by Walborn 1978,  see
also  Figs. \ref{fig:name}, \ref{fig:nameconv}).
Considering also the kinematic data, it thus became clear that
shocks associated with fast-moving visible knots produce much
of the observed diffuse soft X-ray flux.  Since the outer 
knots and filaments have characteristic sizes ranging from a small 
fraction of an arcsecond to several arcseconds, the shock structures
were not well resolved in the ROSAT images.  Seward et al.\ (2001) reported
the first CHANDRA X-ray images in the $0.2-10$ keV band using the Advanced 
CCD Imaging Spectroscopy (ACIS) imaging array with resolution $\approx 
1\arcsec$  
and compared them to HST images.  However, the energy resolution of this early 
CHANDRA observation was  seriously degraded by radiation damage which severely 
hampered the search for spectral variations in the nebulosity.  Here we 
describe a spatially resolved  analysis of the X-ray nebulosity around 
$\eta$ Car, based on a the zeroth-order image from a  deep observation 
obtained using the CHANDRA High Energy Transmission Grating and the ACIS 
spectroscopic array.

\subsection{Aim and advances of the new CHANDRA observations}

We used a dataset\footnote{Dataset No.: 200057 ACIS-S plus HETG; 
P.I.: M.F. Corcoran} which was obtained by the CHANDRA X-ray Observatory primarily to 
obtaine a grating spectrum of the central source. CHANDRA images have 
about 10 times better spatial resolution than ROSAT HRI, allowing
us to better define the X-ray source regions and possible coincidences 
with optical counterparts.  Even though the X-ray resolution is
still a factor of 5 worse than the best optical images, we can
now constrain the identifications fairly well.  In addition to 
morphological comparisons of X-ray, optical and kinematic data (see also
Weis et al.\ 2001), we can also explore the X-ray spectral properties  
of selected regions in the ejecta.
 
Our new data and results are presented in the following way. 
Sect. 2 describes the new X-ray observations and their reduction; 
Sect. 3 briefly summarizes earlier work and data sets which are 
necessary and useful for this paper.  In Section 4 we discuss 
new results from the CHANDRA data. Section 5 contains a
comparison between the X-ray data and the kinematics of the optical gas,  while in
Sect. 6 we present a discussion of our results, and we summarize the main results in Sect. 7.   

\section{CHANDRA Observations}

$\eta$ Carinae was observed on 2000 November 19 with CHANDRA's {\it 
Advanced CCD Imaging Spectrometer (ACIS\/)} and 
{\it High Energy Transmission Grating (HETG\/)}. We used the 0th order 
image situated at the aimpoint on the S3 chip, obtaining spectral 
information from numbers of electrons in photon-detection events 
rather than from the dispersed spectrum.  The spatial resolution was 
about 1\arcsec\ and the exposure time was 91 ksec.  The data were screened 
for bad events, rejecting events graded 1, 5 and 7.  We produced
three separate images representing photon energy ranges 0.2--0.6, 
0.6--1.2, and 1.2--11 keV by using the FSELECT and XSELECT tasks 
in the {\it HEASOFT\/} software package.  Figure \ref{fig:colorchandra} 
shows the CHANDRA
X-ray image in the $0.2-11$ keV band while Figure \ref{fig:images} 
shows the images in the 3 energy bands separately.

Spectra of selected areas were extracted and analyzed using the HEASOFT tasks XSELECT and XSPEC. The 
spectral resolution is roughly 
0.13 keV in the $0.2-1.5$ keV energy range.
First we extracted a spectrum of the entire X-ray nebula, excluding
the central object.  Then we extracted spectra from the five relatively bright regions
shown in Figure \ref{fig:nameconv}, along with the spectrum of the 
central object (which suffers from photon event pileup) for comparison.  
Each of the five selected areas 
represents either an identifiable, coherent large-scale 
structure or an area of special interest 
identified by earlier work, e.g., ``knot\,2'' as
discussed by Weis et al. (2001).  These selections were also 
constrained by the need for good photon statistics.   
Their names in Fig.\ \ref{fig:nameconv} correspond to visual-wavelength features identified by Walborn (1978;  see Fig.\ \ref{fig:name}). 
Each spectrum was corrected for background, using background samples 
close enough to $\eta$ Car so that the intervening absorbing column 
density was approximately constant. Since the diffuse X-ray emission in the Carina H\,{\sc ii} region  varies spatially (e.g.\ Seward \& Chlebowski 1982), we compared spectra from a variety of nearby regions when constructing background spectra.


\section{Supporting work and datasets}

\subsection{HST images}

We compared the X-ray emission to optical images obtained with the
{\it Hubble Space Telescope (HST\/)} and the {\it Wide Field Camera (WFPC\/)}.
Suitable data were retrieved from the STScI archive\footnote{Dataset 
numbers u4460102m...u4460105m from program GO 7253, 
whose PI was J.\ Westphal. }.
In order to map the faint outer knots which are the main contributors 
to the X-ray shocks (Weis et al.\ 2001), we selected four images 
that used WFPC2 filter F658N (see below), exposed for 
2\,$\times$\,200\,s and 2\,$\times$\,4\,s.
The short exposures were used to correct CCD bleeding in the longer
exposures.  Data from all four WFPC2 quadrants were combined as 
a mosaic, with spatial resolution about 0\farcs0996 per pixel  
and a field large enough to include a few stellar X-ray sources, 
which served as astrometric reference points for aligning 
the CHANDRA and HST images.  Figure \ref{fig:hst} shows the  resulting F658N 
image.   Here the display intensity is optimized to show the filamentary 
outer ejecta;  the bipolar Homunculus is saturated in this rendition
but we have marked it with intensity contours.  

Finally, a note concerning F658N, which is usually considered to 
be either a [N\,{\sc ii}] $\lambda$6586 emission line filter or else 
a redshifted H$_{\alpha}$ $\lambda$6565 filter.  Since the outer ejecta 
studied here have a wide range of velocities, both emission lines 
contribute to Fig.\ \ref{fig:hst} with spatially-dependent efficiencies.  
[N\,{\sc ii}] is generally brighter in this gas (Davidson et al.\ 1986),  
but H$_{\alpha}$ dominates the image at locations where the knots 
are sufficiently redshifted, while neither line shows the blueshifted
material well with this filter.  We chose F658N rather than F656N, 
because images taken with the latter have much worse CCD bleeding 
from the central star's intense H$_{\alpha}$ emission.

\begin{figure}
{\resizebox{\hsize}{!}{\includegraphics{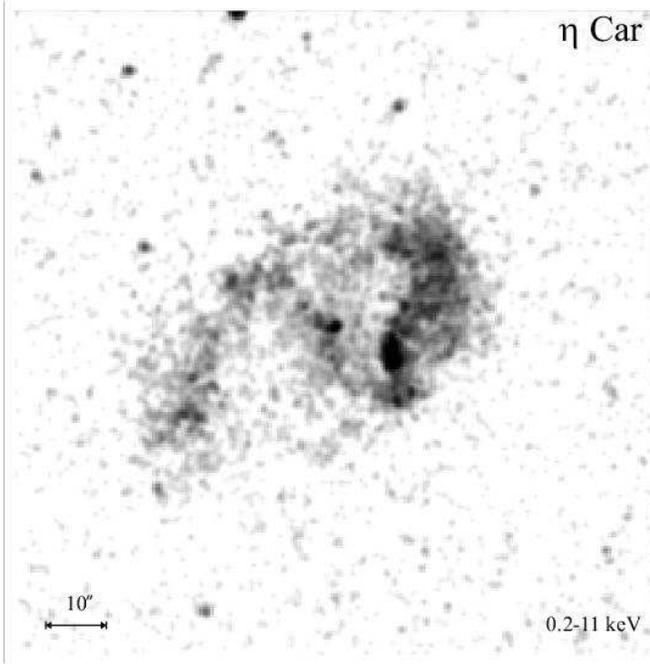}}} 
\caption{
This figure shows a grayscale image of the X-ray emission 
of $\eta$ Carinae's outer ejecta and the star, 
in the energy range between 0.2-11 keV.
The intensity levels are optimized to emphasize the 
softer emission (roughly between 0.5-1.2 keV) from the ejecta.
The bright, hard emission of the central object 
is therefore less pronounced. 
}
\label{fig:colorchandra}
\end{figure}

\begin{figure}
{\resizebox{9cm}{!}{\includegraphics{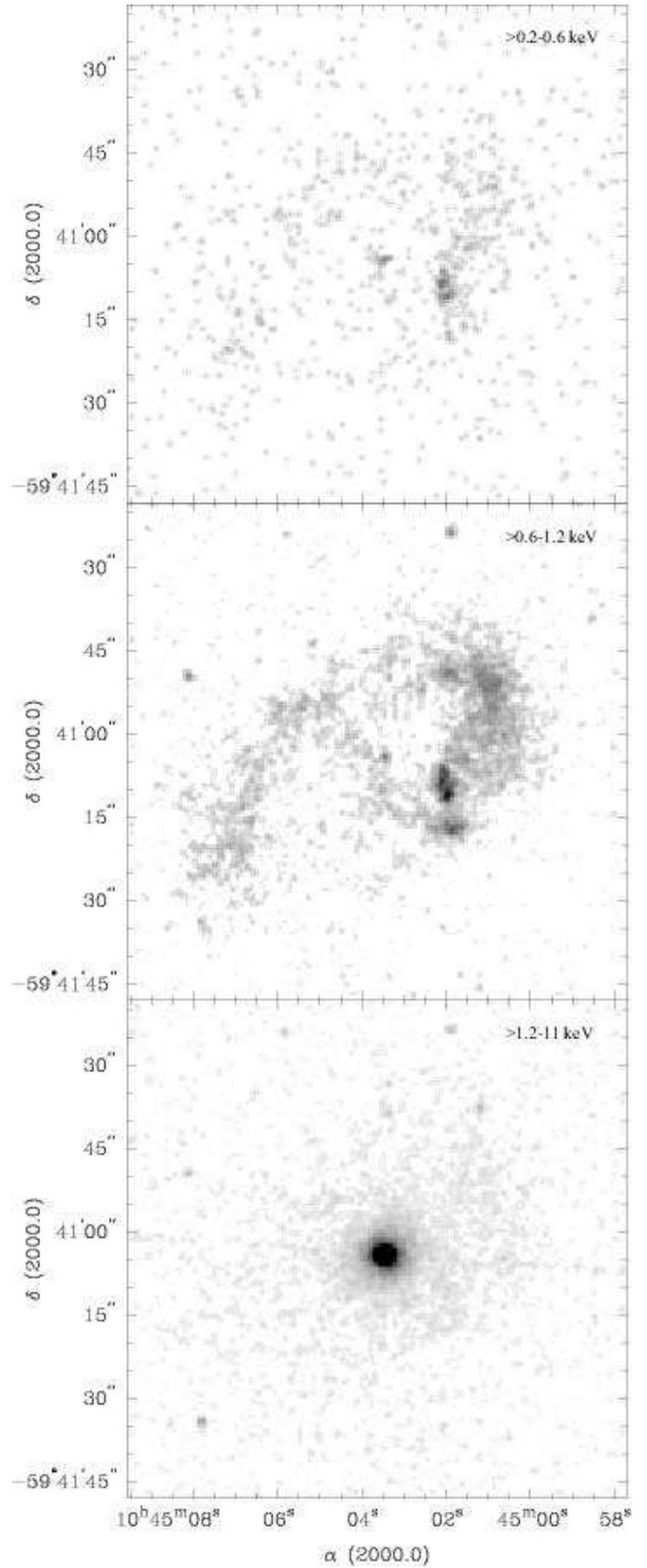}}} 
\caption{ACIS-S3 CHANDRA images of $\eta$ Carinae in
  different energy bands. Upper panel: energy range $0.2-0.6$\,keV,
dominated by the nitrogen K emission line
(see text for details). Middle panel:  energy range $0.6-1.2$\,keV.  
Lower panel: emission at energies $1.2-11$\,keV, where 
the star dominates the emission.
}
\label{fig:images}
\end{figure}

\begin{figure}
{\resizebox{\hsize}{!}{\includegraphics{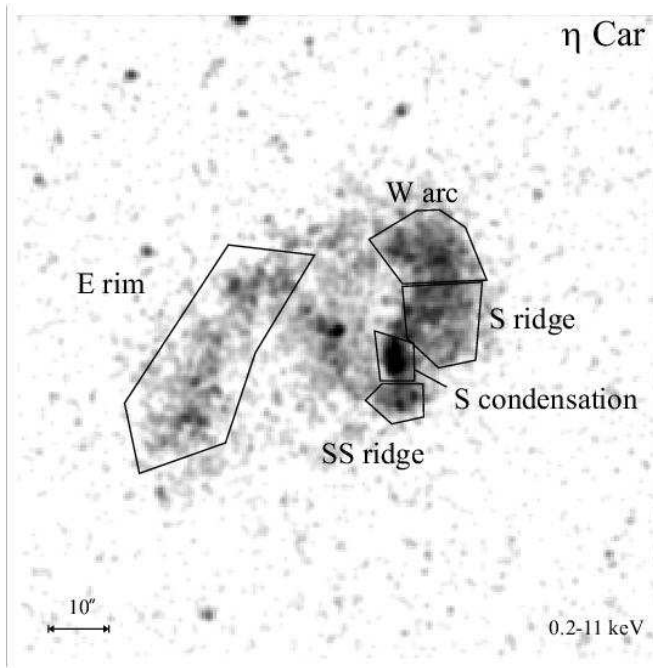}}} 
\caption{CHANDRA full band image
in which the X-ray emitting regions in the outer ejecta are 
subdivided and named to follow the nomenclature used to describe the optical features 
(Walborn et al.\  1978, see also Fig.\ \ref{fig:name}).
}
\label{fig:nameconv}
\end{figure}

\subsection{Echelle Spectra - Kinematics}

We used optical spectra to determine local expansion velocities,
for comparison with both the optical image and the X-ray data. 
This analysis was similar to that reported by Weis et al.\ (2001); in fact 
we used the same kinematic data set. 
Since Weis et al.\ described these spectra in detail, here we 
merely summarize the essential facts.  Data were obtained 
with the Echelle spectrograph of the 
4\,m telescope at {\it Cerro Tololo Interamerican Observatory}.  
The outer ejecta were mapped with 32 slit positions;  the 
slit was oriented at position angle 132\degr\, i.e., parallel to
the major axis of the Homunculus, and successive slit centerlines 
were 1\farcs4 apart.  Severe stray light within the
instrument spoiled six of the central spectra. 
The FWHM spectral resolution was 14\,\kms\ across a spectral range 
of 75\,\AA\ centered on H$_{\alpha}$, while spatial resolution,
limited by seeing,  varied between 1\farcs4 and 2\arcsec\ 
during the observations.

\section{CHANDRA Results}

\subsection{CHANDRA images}\label{sec:images}

\subsubsection{The morphology of the X-ray emission}

We extracted images  in three energy bands from
the ACIS S3 data. The total-band zeroth-order image ($0.2-11$ keV) was smoothed (using a gaussian smoothing with $\sigma = 0.8$) and is displayed in greyscale  in Fig.\ \ref{fig:colorchandra}.    Images in individual energy bands are shown in Fig. \ref{fig:images}. The
softest energy band spans  $0.2-0.6$\,keV. As will be seen 
later from the spectra 
(see section \ref{sec:spect}), the predominant emission  in this band
is from the nitrogen K lines (see next section). 
The intermediate energy band ($0.6-1.2$\,keV, the middle panel in Fig. \ref{fig:images}) will be of principal interest in the upcoming discussion 
since it contains most of the  X-ray emission contributed by the outer
ejecta. The central source is only barely visible in this energy
range. At higher energies ($1.2-11$\,keV, bottom panel, Fig. \ref{fig:images}),
the emission from the 
central source dominates. Since our main interest here is  the
diffuse soft X-ray emission we do not discuss in detail the hard-band emission from the central 
object (see Corcoran et al. 2001b and Pittard \& Corcoran 2002 for a more 
detailed discussion of the X-ray spectral energy distribution of the central 
source). We will however use the images and spectra of the hard
X-ray emission -- and therefore the central object -- to compare with 
the properties of the softer emission of the outer nebula.  Also, since the morphology 
of the X-ray nebula  in the nitrogen K line and at intermediate 
energies are similar, but the 
nitrogen K line image has poor signal-to-noise,
we will concentrate our following discussion on the intermediate-band (0.6 to 
1.2\,keV) images.

The X-ray nebula in the energy range $>0.6-1.2$\,keV (Fig.\ \ref{fig:images} 
middle panel) is roughly hook shaped (sometimes referred to as the ``shell''), as previously seen in the ROSAT images (Weis et al.\ 2001).
Analysis of an early CHANDRA image implied that the X-ray nebula is a ring rather than a limb-brightened elllipsoidal shell (Seward et al.\ 2001)
due to the lack of emission in the center.  
The limited spatial resolution
of the ROSAT images led Weis et al.\ (2001) to attribute the  majority of 
the emission to two 
individual X-ray emission regions which they called ``knot\,1''
and ``knot 2''.  The higher spatial resolution of CHANDRA
separates these knots into several individual components. The region originally 
called knot\,1 contains at least two bright components. The center of 
 knot\,1 is now identified  with the {\it S\,condensation} (see  nomenclature in Fig. \ref{fig:nameconv}), while knot\,1 included (at least
partially) an area we now call the ``{\it SS\,ridge}'' (since it lies just south of the optical feature known as the S\,ridge). 
Combining the S\,condensation and SS\,ridge yields a region of emission which 
matches well the triangular shaped knot\,1 of Weis et al. 2001. 
We identify knot\,2 
with the X-ray bright regions near the {\it W\,arc}
in Fig.\  \ref{fig:nameconv}. 
We identify the X-ray bright region between the W\,arc and the S\,condensation as
the {\it S\,ridge} since it coincides with the northeast extension of the optical S ridge.
Finally, as seen in Fig.\  \ref{fig:nameconv},  we identify a large section of X-ray emission in the east as the ``{\it E\,rim}''.  The southwestern part of the E rim
includes optically identified 
features called the {\it E\,condensations} (see Fig.\ \ref{fig:name}), but the X-ray emission 
stretches further out than these optical emission features.  

The CHANDRA intermediate-band image clearly shows a new feature, a band of diffuse emission that crosses the Homunculus 
connecting the SS\,ridge to the E\,rim just slightly
south of the central point source (see Sect. \ref{sec:bridge}). 
We refer to this feature
as the ``\textit{bridge}''. From ROSAT HRI images, Weis et al.\ (2001) speculated about a possible elongation
of the central source with a position angle of 30\degr. This is 
roughly the same as the 
orientation of the bridge (keeping in mind the lower resolution  
of the ROSAT images). Therefore we conclude that the elongated 
X-ray emission at the position of the central source seen by ROSAT actually was 
a detection of the ``bridge''. The CHANDRA images show that the central X-ray source, however, is consistent with point-like emission.
The intermediate-band CHANDRA image (Fig.\ \ref{fig:images}, middle panel) shows a deficit of X-ray emission in the west just to the north of the ``bridge'', between (clockwise)
the end of the E\,rim and the beginning of the W\,arc.

The $1.2-11$ keV image 
shows nearly exclusively the bright hard central source. As previously noted (Seward et al. 2001) CHANDRA images
(Fig. \ref{fig:images}) seem to indicate
that the central point source is surrounded by a halo of relatively hard emission reaching out into
the W\,arc,  S\,condensation and S\,ridge regions
which is partly 
due to the wings of the point spread function (PSF), 
and partly due to some real hard diffuse emission.
Although the region near $\eta$ Car is badly contaminated by the wings of the point spread function of the bright central source, Fig. \ref{fig:contours} suggests that
some real hard X-ray emission may exist  in the Homunculus, similar to the ``halo''  noted by Seward et al. (2001),  and possibly in the outer nebula as well (see Sect. \ref{sec:hard}).
In order to investigate whether any real hard emission originates in the region around $\eta$ Carinae, we attempted to model the radial brightness distribution of the central source.  We used  
the CHANDRA Interactive Analysis of Observations (CIAO) software package to 
generate a model PSF based on the CHANDRA 
PSF library and the procedure outlined in the CIAO science threads. We chose a central energy of 4.5\,keV to generate the PSF, based 
on the spectrum of the central source (see Fig. \ref{fig:shellstar}), and we used tasks in IRAF/STSDAS for the comparison of the PSF to our data.
We smoothed the model PSF image with the same $\sigma=0.8$ Gaussian 
as the observed hard-band image.  Additionally we corrected the centering of 
the model PSF to agree with the peak position  of the hard source
by fitting Gaussian and Moffat functions.  The 
centering of the two images agreed after this process to better 
than 0.2 pixel.  We then attempted to scale the PSF to the peak flux 
of the central source.  
This scaling is inexact, because the central source suffers from severe photon pile-up which causes the intensity of the central source to be underestimated by a large amount. We experimented with various scalings of the model X-ray PSF to the observed PSF to find the maximum contribution from the wings of the point source.  
We found that even if we adopt a scaling factor of 2,  extended X-ray features in the $1.2-11$ keV image around the central source are untouched by the PSF subtraction and are apparently real.

\subsubsection{Comparing CHANDRA- and HST-images}

\begin{figure}
\begin{center}
{\resizebox{9cm}{!}{\includegraphics{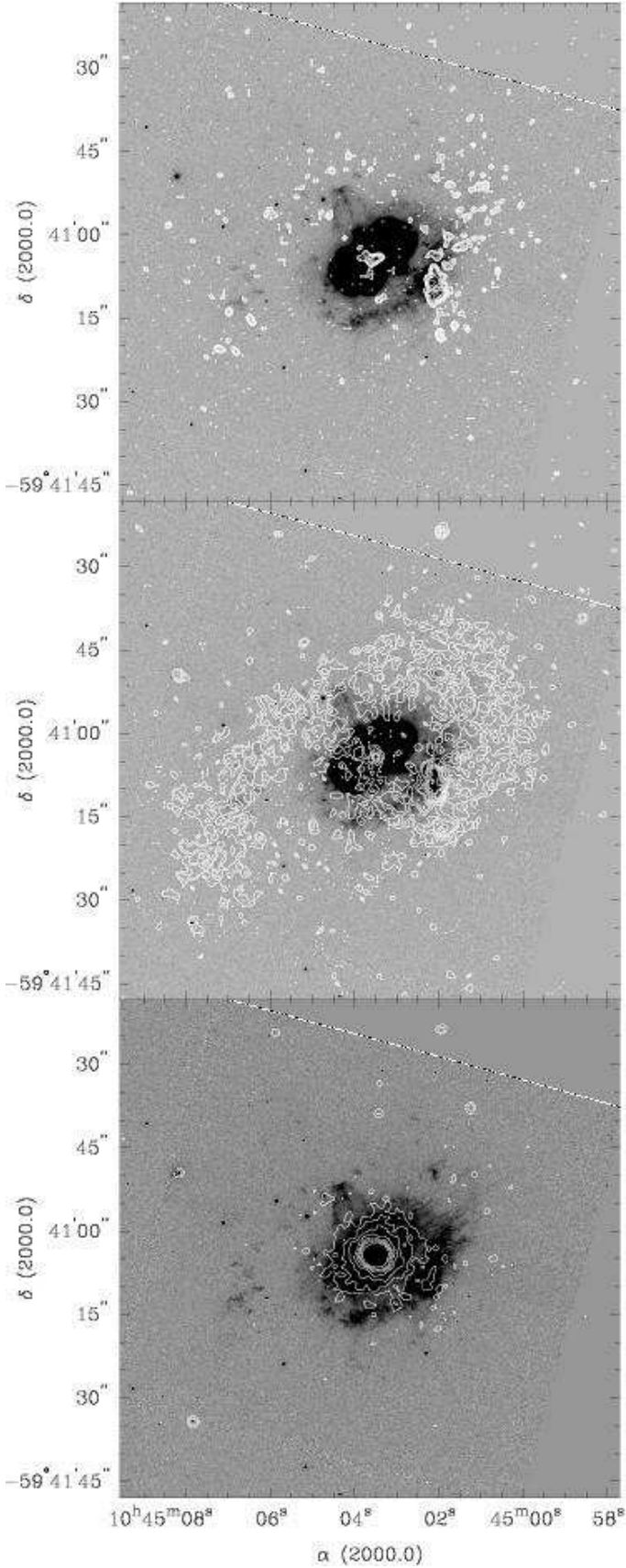}}}
\end{center}
\caption{CHANDRA  contours 
in the same energy bands as in
  in Fig. \ref{fig:images} are compared to an HST F658N image.} 
\label{fig:contours}
\end{figure}

To compare the X-ray emission with the optical emission we contoured the X-ray images in the 3 bands and compared them  with an F658N HST image. All overlays are shown in
Fig. \ref{fig:contours}.
Since it contains most of the emission and therefore yields the best
comparison we will start by comparing the intermediate band CHANDRA image with the
HST optical image. As noted earlier by Weis et al.\ (2001)
only a few X-ray features seem to
coincide with optical features.  The very bright X-ray ``S\,condensation'' 
does 
match the bright optical emission near the S\,ridge, and 
some of the X-ray emission peaks in the W\,arc seem to correspond to 
optical structures, but in general the X-ray and optical emission peaks
differ. Sometimes the locations
are identical but then the shape of the maxima is not. The brightest
optical feature in the W\,arc,  the so-called 
{\it  W\,condensation\/} does show enhanced 
X-ray emission, though it is not the brightest X-ray feature. Here the X-ray emission seems offset from the optical and the
morphology is also slightly different. 
The X-ray emission in the E\,rim coincides with the position of a larger
number of knots in this area, especially  the optical
{\it E\,condensations} (Walborn et al.\ 1978).

Interestingly, an X-ray minimum between the E\,rim and the W\,arc 
agrees well with the trapezoid shaped structure which includes  
the so called {\it N-jet} (e.g. Meaburn et al.\ 1993)
and the  {\it NN} and {\it NS\,condensation} 
(Walborn et al.\ 1978). For simplicity we refer to this 
total trapezoid structure as the {\it NN\,bow\/} (following Morse et 
al.\ 1998, see Fig.\ \ref{fig:name}). This suggests that the {\it NN\,bow\/} lies in front of and absorbs this region of the X-ray ring.

In several cases X-ray emission is found 
in areas where only very faint optical emission
is present. 
This is  especially true for the SS\,ridge and the parts of the W\,arc, the E 
rim (to the south) and the S\,ridge far from the central star.   
However, the large brightness differences
between the fainter outer ejecta and the bright Homunculus 
make it hard to identify optical counterparts to these X-ray structures. 
For the E\,rim some faint optical emission is discernible even at the 
southernmost edge seen in X-rays. At the  outermost parts of
the W\,arc and  S\,ridge (Fig.\ \ref{fig:contours}), however, no 
optical counterparts to the X-ray emitting material were detected.

The upper image in Fig.\ \ref{fig:contours}
shows  the $0.2-0.6$\,keV band and resembles the overlay of the
nitrogen K line image. 
The nitrogen K lines are most prominent
in the S\,condensation, 
parts of the S\,ridge and the W\,arc.
Even though the central source also shows a maximum in this image, this maximum does not result from nitrogen line emission but rather from residuals of the 
continuum emission from the star. 

The hard-band image shows X-ray emission from the 
central object.
A small maximum in the S\,ridge however, turns out to be a
real feature, after subtracting the CHANDRA PSF (Sect. \ref{sec:hard}). 
 
\subsection{CHANDRA Spectra\label{sec:spect}}

\begin{figure}
{\resizebox{\hsize}{!}{\includegraphics{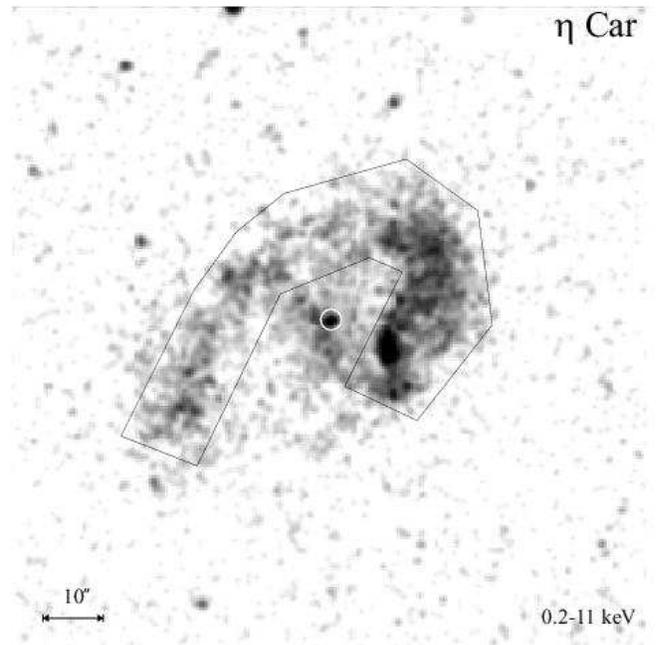}}} 
\caption{The CHANDRA image with the spectral extraction regions shown.
}\label{fig:posshellstar}
\end{figure}

We extracted one spectrum from the entire ``hook'' (see
Fig.\ \ref{fig:posshellstar})
excluding the central source, and  one spectrum for each region defined in  
Fig.\ \ref{fig:nameconv}. For comparison a spectrum of the central source was
also extracted. The extraction area for the central source
(see Fig.\ \ref{fig:posshellstar}) was chosen
to be as small as possible, to avoid overlap with the outer ejecta, 
but at the same time large enough to obtain a reasonable signal to noise
ratio and to include most of the flux from the point source.   Because of this, and because of photon pile-up, the flux of the 
central source is therefore underestimated.

\begin{figure}
{\resizebox{\hsize}{!}{\includegraphics{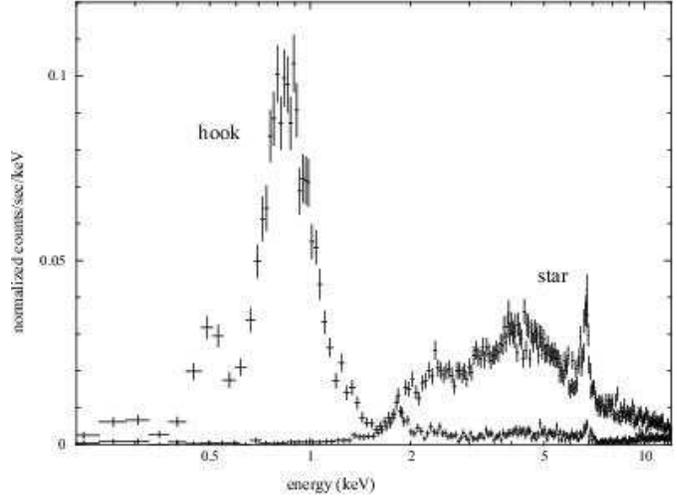}}} 
\caption{The X-ray spectrum of the ``hook'' compared with the 
spectrum extracted from the star. The total flux of the star is  actually much higher than shown, since emission from only a small central 
region has been included in the extraction, and due to photon pile-up in the detector.}\label{fig:shellstar}
\end{figure}

Fig.\ \ref{fig:shellstar} shows the spectrum of the entire X-ray nebula and, 
for comparison, the spectrum of the central source. 
Because the temperature and compositions vary through the X-ray nebula,  
we did not attempt to model this spectrum.

Fig.\ \ref{fig:shellstar} shows the general features of the X-ray spectrum of
the outer ejecta. The strongest emission is between 0.5 to 2 keV.
The emission is mainly thermal -- see also discussion later in the text -- but
only a few lines/line complexes are visible in this region. 
Between  $0.4-0.55$\,keV the larger bump, clearly separated from the continuous
emission between 0.55 to 2 keV, is emission from the  nitrogen K
lines, as first noted in an ASCA observation by Tsuboi et al. (1997).   The major contribution is 
from the 
N\,{\sc vii} Ly$_{\alpha}$ lines at 0.50 and 0.49\,keV 
(24.7792 and 24.7846\,\AA). 
A smaller fraction results from various  N\,{\sc vi} He$_{\alpha}$ lines 
around 0.43\,keV (28.7870; 29.0843 and 29.5347\,\AA; see also Leutenegger 
et al.\ 2003). The high intensity of this blend of nitrogen 
lines indicates a nitrogen overabundance. 
This is consitent with the nebula of $\eta$ Carinae being 
CNO-processed material -- which
manifests itself in a higher nitrogen abundance (accompanied by oxygen and
carbon underabundance, e.g. Davidson et al. 1982).
In the softest region a second peak occurs around 0.3\,keV 
which we tentatively identify with the Si\,{\sc xii} Li$_{\alpha}$ lines at 
(40.9110 and 40.9510\,\AA). However, since the sensitivity and calibration
is inferior in this energy band, and the resolution 
too low to seperate individual lines, this detection needs further 
confirmation.
The spectrum of the hook also shows 
a peak around  1.8\,keV, which is a blend of various Si-lines:
Si\,{\sc xiii} He$_{\alpha}$ (6.6479\,\AA) and Si\,{\sc xiv} H$_{\alpha}$
(6.1804 and 6.1858\,\AA). Very likely also the Mg\,{\sc xii} H$_{\alpha}$
(7.1058 and 7.1069\,\AA) lines contribute here. 
Above 2 keV, emission from the hook is barely visible. 
A more detailed discussion of the hard spectrum of the outer ejecta
follows in Sect. \ref{sec:hard}.

The stellar spectrum itself is much harder and all emission below 
1.8\,keV  is absorbed by the Homunculus nebula. Since the star is embedded in
the dense nebula soft emission cannot penetrate to us, only the soft
emission from the outer ejecta can reach us. Below 0.4\,keV
X-ray emission will also be heavily absorbed by the forground H\,{\sc i}. 
The spectrum of the star is dominated by hard emission and
the Fe K line complex. 

\begin{figure}
{\resizebox{7.5cm}{!}{\includegraphics{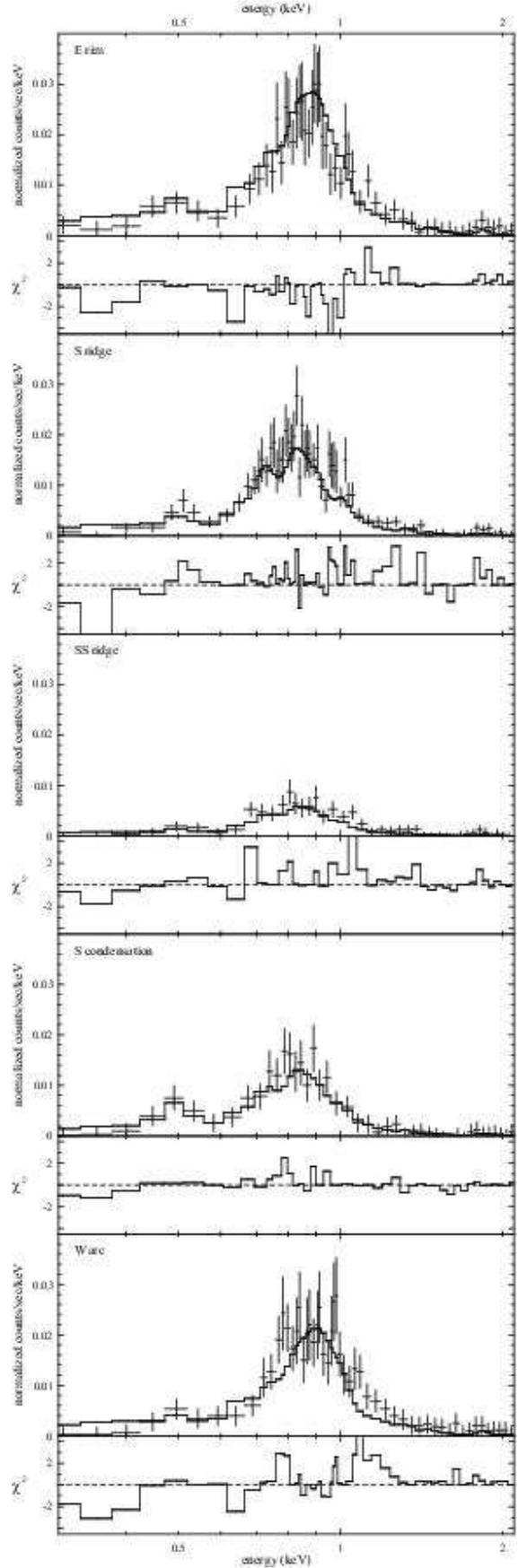}}} 
\caption{X-ray spectra extracted from  select regions shown in Fig. \ref{fig:nameconv}). A 
collisionally ionized Mewe-Kaastra-Liedahl plasma emission model has been 
fitted to all spectra.  We assume solar
abundances for all elements except for  nitrogen. 
}
\label{fig:spectra}
\end{figure}

Individual spectra for all 
areas defined in Fig.\ \ref{fig:nameconv} have been extracted
and are shown in Fig.\ \ref{fig:spectra}. 
These plots
range from 0.3 to 2\,keV and the scale is again logarithmic. 
The range of the vertical axes are the same in each plot to visualize more easily intensity differences 
between each region. 
Fig.\ \ref{fig:spectra} also includes
the model fits, which will be discussed later. 
Underneath each spectra the corresponding 
plot of the $\chi^2$ deviations of the model is indicated. 
We can seen that except for the intensity, the spectra are quite similar 
to each other and do not differ much from the overall spectrum of the total 
outer ejecta.
Most of the X-ray emission lies between 0.6 and 1.2\,keV. 
The emission from the nitrogen lines varies spatially, implying abundance 
variations by a factor of 5 through the outer emission.

Each spectrum was 
fitted using the XSPEC analysis package and a Mewe-Kaastra-Liedahl thermal plasma model (e.g.\ Mewe et al.\
1985, 1986) using the XSPEC spectral fitting software. This model includes line 
emissions from several elements and assumes  coronal equilibrium.
We used a foreground hydrogen column density of $\log N_{\rm H}$=21.3 (e.g.\ 
Savage et al.\ 1977). 
We adopted solar abundances for all elements with the exception of nitrogen, 
and kept the N abundance, temperature and emission measure as free 
parameters.  
We checked our model fits not only by inspection and minimization of 
the $\chi^2$ values but  by  comparing various other 
model fits to the data. Fits using Raymond-Smith plasma
models yielded qualitatively  acceptable results as well, though quantitatively,
the deviations of the models from the spectra were somewhat larger.
Bremsstrahlung models failed since no lines are included. This might also 
explain the large difference
between our results and the temperature derived by Seward et al.\ (2001).
Finally we also tested  Mewe-Kaastra Liedahl models with a purely
solar abundance for all elements and did not obtain any good fits.

Table \ref{table1} summarizes the results from each model fit to the spectra
for each  region. Temperatures are given in keV and K,
along with the nitrogen abundance relative to solar abundance. 
Note that for the abundances we use the XSPEC convention which defines
the abundance of a particular element as  the number of nuclei per 
Hydrogen nucleus relative to solar abundances (Arnaud \& Dorman 2002).
In the last column the velocity of the shocking gas
for the corresponding plasma temperatures are plotted.
The shock velocities and gas
temperature are related as T = v$^2$ (3$\mu$/16k) 
where v:= velocity of the shocking gas, T := post-shock 
temperature, k := Boltzman constant and $\mu$ := molecular weight
(e.g.\ Landau \& Lifschitz 1966, McKee 1987). 
With the temperature in keV one derives 
that v = 710$\sqrt{{\rm T}_{{\rm [keV]}}/\mu}$.
The velocities in Table \ref{table1}
have been calculated for different assumed values of the molecular weight: 
$\mu$=0.59 is for a completely ionized primordial 
hydrogen-helium mix of X=0.76 and Y=0.24, while $\mu$=0.67 results from a fully ionized gas with a mixture of 
X=0.6, Y=0.4. The contribution of higher metals can be neglected since they have
no significant influence on the molecular weight. 
The latter ratio is in accordance with  
values derived for $\eta$ Carinae's outer ejecta (Davidson et al.\ 1986). 
Table \ref{table1} shows that the velocities of the shocking gas for such a
mixture of heavier and processed elements (predominantly more Helium) are 
reduced  -- roughly by
45\,\kms\ -- compared to the velocities for a primordial gas mix. 
Both values for $\mu$ are given in this context to demonstrate the dependence 
of the post-shock temperatures on molecular weight. This is of interest, in particular, 
since theoretical models (Garc{\'\i}a-Segura et al.\ 1996) 
for LBV nebulae predict values for the He abundance up to
Y=0.76. This would increase the value of the molecular weight to as high as 
$\mu$=0.9 and similarly decrease the inferred velocities.  

Table \ref{table1} shows that the W\,arc is  
the hottest region, followed by the E\,rim. The S\,ridge, SS\,ridge
and S\,condensation show similar temperatures with the S\,ridge being slightly
cooler. The nitrogen abundance is about 6 times solar on average if we exclude 
the S\,condensation, which has a derived nitrogen abundance of 22 relative to 
solar.

\begin{table*}
\begin{center}
\caption{Parameters of the models fitted to the individual spectra}
\label{table1}
\begin{tabular}{lccccc}
\hline
Area & Temperature (keV) & Temperature (K) & Nitrogen  &
V (\kms) / $\mu$=0.59 & V (\kms) / $\mu$=0.67\\
\hline
\hline
E\,rim & 0.70$\pm$0.03\,keV & 8.0$\pm$0.4 10$^6$\,K & 8 & 773\,\kms 
& 726\,\kms \\
S\,ridge & 0.60$\pm$0.02\,keV & 7.0$\pm$0.2 10$^6$\,K & 6 & 710\,\kms 
& 670\,\kms\\
SS\,ridge & 0.63$\pm$0.04\,keV & 7.3$\pm$0.5 10$^6$\,K & 6 & 735\,\kms 
& 690\,\kms\\
S\,condensation & 0.63$\pm$0.04\,keV &  7.3$\pm$0.5 10$^6$\,K & 22 
& 735\,\kms & 690\,\kms\\
W\,arc & 0.76$\pm$0.03\,keV & 8.8$\pm$0.4 10$^6$\,K & 4 & 807\,\kms 
& 756\,\kms \\
\hline
\end{tabular}
\end{center}
\end{table*}

\section{Comparison of CHANDRA spectra and images with the kinematic data}

\begin{figure*}
\begin{center}
{\resizebox{13cm}{!}{\includegraphics{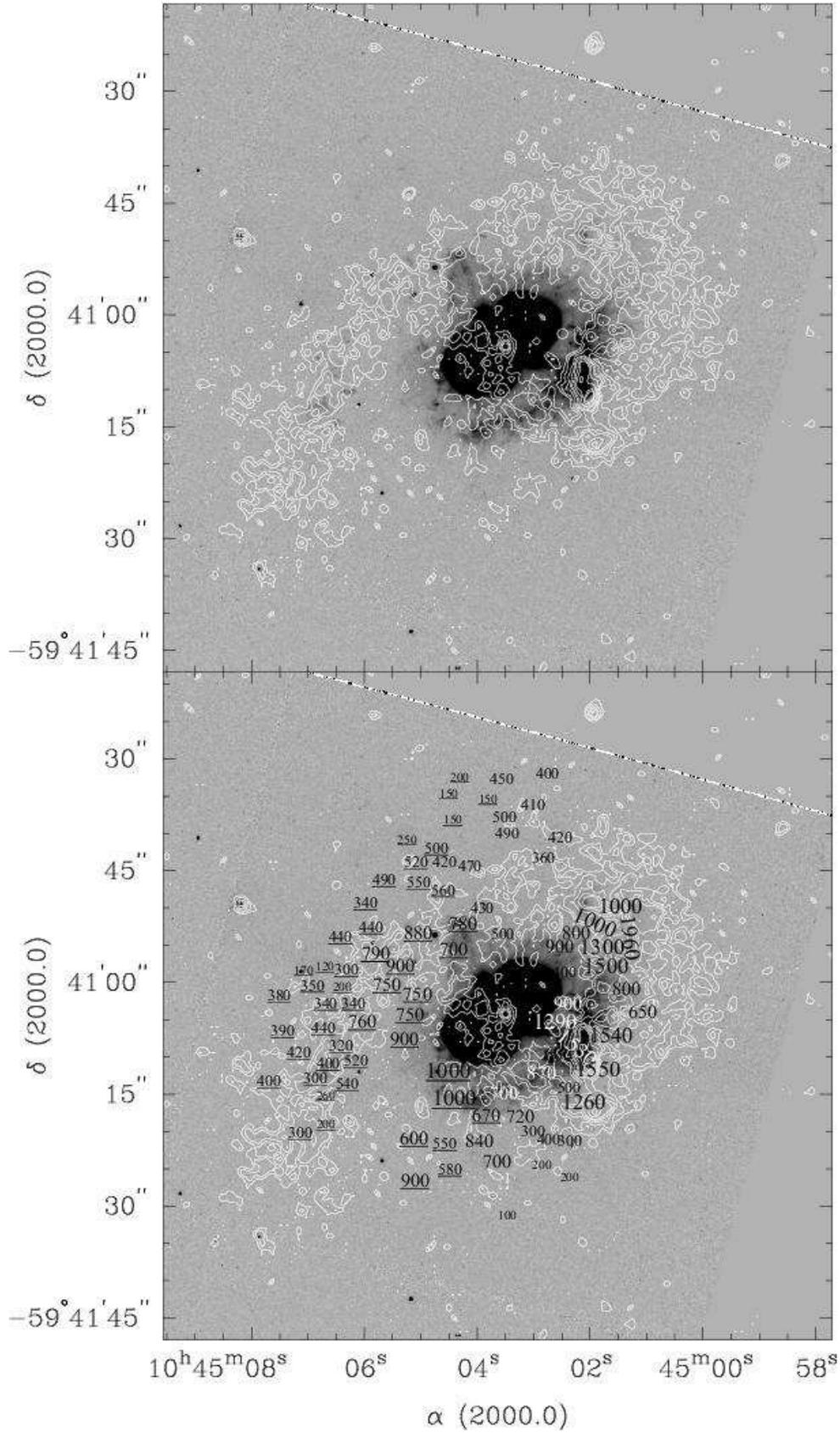}}}
\end{center}
\caption{An overlay of the CHANDRA 0.6-1.2 keV image contours and 
our radial velocity measurements made with the 4\,m Echelle Spectrograph on the optical F658N HST image 
(gray scale).
} \label{fig:overlay}
\end{figure*}

As shown by Weis et al.\ (2001), the soft X-ray emission from the outer ejecta
of $\eta$ Carinae is produced by shocks from fast moving structures.
They showed that X-ray maxima, knots\,1 \& 2,
correspond to  regions where the fastest optical features are also located.
In the following we compare the kinematic data with the 
CHANDRA images and spectra of each analogous region. In 
Fig.\  \ref{fig:overlay}
the upper panel displays the contours of the CHANDRA image ($0.6-1.2$ keV)
on the HST image and in the lower panel velocities in \kms\ have
been added. Blueshifted velocities are underlined and the 
font size of the characters increases with increasing 
velocities. With the higher resolution of
the CHANDRA images we find in many cases an even better agreement between 
X-ray brightness and velocity.
The brightest X-ray structures still are those with the highest velocities.
The S\,condensation, for e.g., has velocities of 1290 to 1550\,\kms\  
and in the SS\,ridge we find a 1260\,\kms\ feature. In the area of the W\,arc 
the expansion velocities are up to 1960\,\kms.  
Note that Fig.\ \ref{fig:overlay} shows only the maximum velocities
of each region for clarity; in reality each area actually 
spans a range of velocities (see Sect. \ref{sec:plasma}). 

For the best prediction of the X-ray luminosity and temperature, we have to weight the measured velocities with the densities of 
the gas. This is unfortunately not possible in a quantitative way using 
the present data.  The echelle spectra do not include density sensitive 
lines, and an estimate from the H$_{\alpha}$ surface brightness (using 
emission measure and assuming a geometry) is not possible due to the 
overlap of the strong red- and blue-shifted [N\,{\sc ii}] lines with 
H$_{\alpha}$ in the H$_{\alpha}$ filter band. Only few [S\,{\sc ii}] based 
density estimates in the outer ejecta have been reported 
(Dufour et al.\ 1997, Weis 2002) 
and all are near the high density limit of the line diagnostic.  
A very rough qualitative 
estimate based on the surface brightness of the H$_{\alpha}$ line in the 
echelle spectra gives about 800\,\kms.  This is a typical gas velocity 
in the regions where we find the highest peak velocities (like the S\,ridge or W\,arc), 
which are often those having a very broad velocity distribution (up to at least 500\,\kms; 
Weis \& Duschl, in prep.).  This raises the possibility of clump interactions.
In most of the cases the highest velocity in a certain area 
implies a higher mean velocity in that region.

Due to stray light contamination from the bright central source, 
we cannot constrain the expansion
velocities for the ``bridge'' region across the Homunculus.
Also  no measurements are
available for the southern part of the E\,rim. However, one usable spectrum from this region indicated that 
expansion velocities as high as 1000\,\kms\ are present. This data-point
has not been included on Fig.\  \ref{fig:overlay} since more
measurements are needed for confirmation.

The shock velocities derived from the X-ray  temperatures are listed in 
Table \ref{table1}.
Comparison with the radial velocity field based on our Echelle mapping 
shows that the typical velocities of the majority of the clumps 
are in agreement with the postshock temperatures from the fit to the X-ray  
spectra. 
The expansion velocities of the outer ejecta 
are therefore sufficiently high to explain the soft diffuse X-ray emission
and the derived temperature of the gas. 

The range of velocities detected in the outer ejecta field gives a natural 
explanation for the complexity of the X-ray spectra.
In the areas which contain the highest velocities the velocity distribution
of the clumps is generally also shifted to higher median velocities.
The area, for instance, which contains the fastest expansion
velocities (v$\sim$1900\,\kms) -- the W\,arc -- shows 
the hardest X-ray spectrum (T=8.8 10$^6$\,K). 
Most of the clumps in this region have
velocities between 700-900\,\kms. These expansion velocities are 
in good agreement with those expected for a gas with that temperature 
(see Table \ref{table1}). 

The E\,rim also shows a relatively high temperature,  T=8 10$^6$\,K,
but the radial velocities are on average lower than in the W\,arc. 
However, the velocities detected (see Fig.\ \ref{fig:overlay})
and the velocities derived from the X-ray spectra (see Table \ref{table1}) are in  good 
agreement. We also have to keep in mind that we measure 
only radial velocities so that the real expansion velocities are
larger.
 For instance this is the case with the NN\,condensation and the NS
condensation (which are actually part of the  E\,rim as defined here). These
structures show a radial velocity of about 700\,\kms\ but the tangential 
velocities are 1360\,\kms\ and 1180\,\kms\ for the NN and the NS\,condensation
respectively (Walborn et al.\ 1978).

\section{Discussions and conclusions}

\subsection{Origin of the X-ray nebula's shape}

The observed properties of the X-ray emission is a combination 
of  the speed of the outflowing material and the density of the ambient medium 
into which the material is moving.  The density of the ambient medium has to 
be sufficiently high to produce detectable emission.

Velocities of the warm gas are derived from our Echelle data supplemented by 
transverse velocities from proper 
motion measurements (Walborn et al.\ 1978).  Densities of the ambient medium 
are derived from direct measurements using the [S\,{\sc ii}] lines 
(Dufour et al.\ 1997, Meaburn et al.\ 1993), but for most of the gas 
we have to rely on 
emission measure derived from HST H$_{\alpha}$ images, which can be contaminated by 
red- or blueshifted [N\,{\sc ii}] as previously discussed.
While the absolute values for n$_e^2$ is somewhat compromised by such 
contamination, the relative surface brightnesses still trace the relative 
overall column densities of the H\,{\sc ii} gas.  

With this  information we have to understand why there is no strict 
correlation between the warm gas seen in the optical and the hot X-ray gas.
There are two alternatives which might explain 
why the density is higher at the places which are  X-ray bright. 

The first possibility is an enhancement of  density produced by an older stellar wind.
During its main sequence phase, $\eta$ Carinae would have possessed a fast, low density wind which would  have formed an interstellar  bubble (Weaver et
al.\ 1977). The interior of this bubble would be nearly evacuated  since all material inside the bubble would be
swept up to a  radius of the order of 70\,pc .   
As $\eta$ Carinae evolved into a blue supergiant and LBV, however,  the wind velocity would decrease 
(down to several 100\,\kms)  and the mass loss rate would increase 
to 10$^{-3}M_{\sun}yr^{-1}$ (e.g. Langer et 
al.\ 1994, Garc{\'\i}a-Segura et al.\ 1996), which would increasingly fill the 
bubble with denser gas. After the ``Great Eruption'' the expansion of the outer 
ejecta  into the remnants of this older, denser LBV wind (which at 0.35\,pc has a density of roughly 150 times that of the ISM)
would be able to slow down the ejecta to its current 
expansion velocity (Weis \& Duschl, in prep) and could produce sufficient X-ray emission in the outer region of the nebula. Had this stellar wind been non-spherical, the ejecta would encounter the  denser wind at different distances from the star.

Alternatively, higher densities may manifest localized regions in which the outer
ejecta consists of many turbulent clumps which have different velocities.  These clumps would collide into and over take each other, so the density is locally 
enhanced. 
This seems a reasonable explanation especially in the S\,condensation where a 
large number of clumps with different velocities are detected. 
X-ray emission is formed in regions with many clumps interacting with high
velocities. In the E\,rim and outer regions of the S\,ridge fewer knots are
detected and one might speculate that here the outer ejecta collides 
into the older stellar wind.  

The Homunculus nebula itself is expanding at a velocity of 650\,\kms\ on 
average but is not visible in soft X-rays, though velocities this large 
should produce soft X-ray emission. The lack of detectable soft emission 
from the Homunculus is likely due to the low density (as judged from the 
low surface brightness in the H$_{\alpha}$ and N\,[{\sc ii}] HST images) 
in the medium into which the Homunculus is  expanding, which has been 
cleared out by earlier episondes of mass loss. 
Therefore the Homunculus expands 
practically freely and no shocks are formed.

\subsection{Velocity fields and plasma temperatures\label{sec:plasma}}

The ACIS-S spectra of the outer ejecta are decently fitted with 
single temperature coronal equilibrium models as we showed in 
sect. \ref{sec:spect}.  Still, looking at the $\chi^2$ deviations, 
residual spikes are 
visible which indicate poorly fitted lines/line complexes.  Indeed, 
several of these spikes correspond to lines complexes of e.g. 
Fe\,{\sc xvii} and Fe\,{\sc xviii} around 0.8 keV, 
Ne\,{\sc x} and Ne\,{\sc ix} at about 1.0 keV, Mg\,{\sc xi} and  Mg\,{\sc xii}
at about 1.3 keV and finally Si\,{\sc xiii} and Si\,{\sc xiv} at around 
1.8 keV. 
While Doppler line velocity displacements are 
big enough to create noticeable shifts in several places in the outer ejecta , 
the highest velocity regions only represent a minor 
part of the total emission of the outer ejecta as seen in our Echelle 
spectra.  Additionally, significant residuals are also visible in 
the X-ray spectra of regions without extremely fast moving knots.
Therefore velocity offsets cannot fully explain the residuals.
As traced by the Echelle data, we see not one typical 
expansion velocity but a distribution of velocities in all regions 
of the outer ejecta. Therfore we have different  
velocities which imply that the X-ray emission is composed of a 
distribution of emissivities, so that our single temperature fits are 
too simplistic.  

The plasma properties we derive are therefore not the properties of the 
whole region, but rather the density weighted properties. We 
see the dominant emission component of a complex plasma, which is most 
probably also the reason why our single temperature fits are relatively good  
despite the multi-component and, due to the short timescales, 
non-equilibrium nature of the X-ray plasma in the outer ejecta 
(Weis et al.\ 2001). 
Our interpretation of the optical gas velocities as shock velocities 
rest on the details of the mass loss history of the LBV. The 
ejecta do not expand into a fast freely expanding wind as 
assumed by Leutenegger et al.\ (2003), but a dense very slow 
moving wind as explained above.  The temperatures of the X-ray 
emitting gas are in good agreement with the velocity of the dominant 
emission component (weighted by surface brightness, therefore 
column density).  What was 
missing is the very hot gas corresponding to the extremes of the 
velocities.  This we may have found now with the ACIS  data.

\subsection{Missing X-ray emission from the NN\,bow}

We detected soft X-ray emission from all the optical line 
emitting regions outside the Homunculus and the presumably low density 
envelope around it. 
Besides this region, there is one notable structure without X-ray emission: 
the NN\,bow. 
This roughly trapezoidal region is regarded as part of the outer ejecta and 
appears as a possible radial extension 
of the equatorial disk (Duschl et al.\ 1995). Surprisingly, the 
region is devoid of all soft emission, as can best be seen in 
the overlay of the soft X-ray emission on the HST WFPC2 F658N image
(Fig. \ref{fig:overlay}, middle panel).
Since the radial velocities of the ionized gas belonging to the NN\,bow 
are typically around $-650$\,\kms\ (Meaburn et al.\ 1993,  Weis et al.\ 2001) 
(similar to other regions 
of the outer ejecta), one would expect to see similar X-ray emission.  
The apparent minimum in the X-ray emission around the
NN\,bow could be explained as a particularly deep X-ray shadow. 
The soft X-ray flux limit is as low as the one observed for 
the Homunculus itself.  Assuming again shocks as the creation mechanism of 
the X-ray emission we would expect gas with a plasma temperature around 
0.56\,keV, and the total absence would point at an H\,{\sc i} foreground 
density of about 10$^{22}$ cm$^{-1}$, similar to the column density through the Homunculus.
Still, since the gas in the NN\,bow is blueshifted, one would expect to 
see this gas on the front side of the dense gas. It appears therefore that the NN\,bow
is expanding into a low density region. Nothing is stopping the expansion in 
line of
sight, the density of the ambient medium is too low to create soft X-rays, 
and there are 
also -- in contrast to the S\,condensation -- no faster knots or clumps 
overtaking the slower gas in this region. 
At the northern end, however, shocks might form, and
here at the optically bright NN\,condensation X-ray 
emission is present. It is further possible that X-ray emission behind the 
NN\,bow is  absorbed by
the NN\,bow itself.  The only measurements of the density of the gas in 
the NN\,bow is  a lower limit from [S\,{\sc ii}] line 
ratios which as mentioned above are close to the high density 
limit (Dufour et al.\ 1997, Weis 2002).  The 
density of the outer ejecta (outside the bright clumps) appears rather 
homogeneous (Weis et al., in prep.) with at least  
n$_e \sim 10^4$ e$^-$ cm$^{-3}$,
consistent with the NN\,bow lower limit.  For an assumed cylindrical geometry
and size based on the observed dimensions in the HST images (diameter of the
cylinder d=0.065\,pc) we get therefore an absorbing 
column of at least $\sim 2 \times 10^{22}$ cm$^{-2}$ through the NN\,bow.
This is large enough to completely hide any $0.6$ keV plasma 
at the sensitivity of our data.

Another interpretation is that an X-ray shadow could be formed if  the 
equatorial disk (Duschl et al.\ 1995) or its extension shadows 
the soft X-ray emission in the center and western part of the NN\,bow.
In this scenario the expansion of the disk produces X-ray emitting 
shocks only at the outer edges, forming the bright X-ray rim at the NN\,condensation and 
the ``bridge'' (see also next section).
Unfortunately, with the present data set details of the absorption 
and possible changes of the absorption across the NN\,bow cannot be 
derived due to a lack of X-ray photons.
It is of interest to note that there is an X-ray bright spur 
just on the opposite side of the NN\,bow (the S\,condensation) which is 
spatially consistent with the plane of the equatorial disk.

Thus, both self absorption and/or expansion into low 
density environment can explain the lack of soft X-rays from the 
NN\,bow.  Much deeper X-ray observations are needed to sort out the 
individual contributions of the processes.

\subsection{The X-ray bridge across the Homunculus\label{sec:bridge}}

In addition to the general lack of soft X-ray emission from the Homunculus 
region we can see in the CHANDRA data a ``bridge'' of soft emission 
crossing the Homunculus from the south-east to the north-west (see Fig.\ 
\ref{fig:colorchandra} and Fig. \ref{fig:overlay}).
We extracted a spectrum from this structure and find that it is
rather similar to the X-ray spectra of the outer ejecta. A single 
temperature Mewe-Kaastra-Liedahl model as fitted to the other spectra (solar
abundance except for nitrogen) gives an adequate fit to the data 
and yields a temperature of 0.67\,keV and a nitrogen overabundance 
of a factor of 4. This is in good agreement with values found in other regions
of the outer ejecta.
The bridge has a somewhat lower X-ray surface brightness than most of 
the other regions in the outer ejecta (middle panel
Fig.\ \ref{fig:images}), but shows similar small scale 
variations in the surface brightness.  It appears from all the properties 
as another part of the outer X-ray nebula.

The most straightforward interpretation for the soft X-ray bridge across 
the Homunculus is therefore a clump or a group of clumps located physically 
in front of the Homunculus, which was not seen in previous observations due 
to insufficient spatial and spectral resolution. 
The bridge is slightly below the optical disk and might represent 
shocks at the very outer ends of the disk, which in projection 
would appear offset from the bright optical emission of the disk. 
Unfortunately, the strong scattered light from the Homunculus prevents us 
from detecting related H$_{\alpha}$ emitting gas in our Echelle spectra and 
therefore determining the velocity field of this gas.
In the HST images, any optical extension of the disk is also difficult
to detect due to the very bright optical emission of the Homunculus.

\subsection{Hard X-ray emission from the nebula around $\eta$ 
Carinae\label{sec:hard}}

We have shown in section \ref{sec:images} 
that some of the hard X-ray emission coinciding with the S\,condensation and 
the SS\,ridge are untouched by the PSF subtraction and therefore real.  
This implies that the S\,condensation and some of the SS\,ridge 
show hard emission, consistent with plasma temperatures comparable 
to the central source.  The implied shock velocities are well in 
excess of 1500\,\kms.
Unfortunately the number of photons in both features are too small to 
generate a meaningful spectrum, but the result lends some support 
for the notion that not all of the hard tail and the Fe\,{\sc xxv} 
visible in the total nebular spectrum (Fig.\, \ref{fig:shellstar}) 
are due to wings of the central 
source's PSF, but intrinsic to the highest velocity knots in the 
nebula.\\

\subsection{No X-rays from the strings}
Some of the most puzzling features of the outer nebula are 
several extremely narrow optical emission filaments, dubbed ``strings'' 
(Weis et al. 1999), ``whiskers'' (Morse et al.\ 1998) or 
``spikes'' (Meaburn et al.\ 1996).
Because of their high velocity -- nearly 1000\,\kms\ (Weis et al. 1999) --
and large density -- n$_e \sim 10^4$ e$^-$ cm$^{-3}$ (Weis 2002) it 
is worthwhile to search for X-ray emission from these structures. 
Detection of such emission would constrain the properties of the interaction 
of the ``strings'' with the surrounding 
medium and would  further limit the possible creation 
mechanisms of these enigmatic features. 
The northwestern strings (\#3 and  \#4 as designated in Weis et al. 1999) are located in the complex region 
of the W\,arc. Plenty of diffuse X-ray emission is present, but no emission 
can be convincingly associated with the strings. 
The southeastern  strings are located in a much less crowded region.  
Unfortunately no diffuse X-ray emission can be seen either at the tip or 
along the body of the strings (\#1, \#2 and \#5).  
The reason for this non-detection is most probably in the very small 
working surface of the strings (Weis et al., in prep.).  Much deeper 
CHANDRA data will be needed to search for the interaction of the 
strings with the surrounding material.

\section{Summary}

With the new CHANDRA data we can conclude that the hook shape of the 
X-ray emission is not solely determined by the distribution of fast 
moving ejecta, as implied by Weis et al.\ (2001), but  also by absorption of 
the soft X-rays and by the distribution of denser gas far from the
Homunculus. The density enhancement could either be 
due to a remnant stellar wind and/or -- as seen 
from the HST images and 
kinematics -- from the mutual interaction of the outer ejecta with itself, 
which would yield the more patchy nebular structure due to the 
non-uniform distribution of the fast and slower moving gas knots. 
In general the detected velocities of the majority of the clumps in the outer
ejecta are consistent with the velocities expected from 
the X-ray gas with temperature derived from the CHANDRA spectra, 
supporting the presence of a low velocity, dense LBV wind 
before the ``Great Eruption''.
This also naturally explains the missing X-ray emission of the expanding 
Homunculus.
The surprising absence of diffuse X-rays from the NN\,bow can be explained 
by its  nearly freely expansion in line of sight. Only at the 
northern tip -- the ``NN\,condensation'' -- an X-ray bright spot is visible; 
this may be indication of shocked emission. 
As a second possibility the NN\,bow could  be the outer part of a 
dense expanding equatorial disk as proposed by Duschl et al.\ (1995) 
which shadows parts
of the NN\,bow in X-rays so only the rim of the disk which forms shocks is
visible in X-rays as the bridge and the bright rim of the NN\,bow. 

The X-ray spectra are well approximated by single temperature collisional 
equilibrium models assuming solar abundances for all elements except nitrogen. 
The N abundance is strongly enhanced, consistent with the optical 
spectra (Dufour et al.\ 1997), earlier ASCA (Tsuboi et al. 1997), CHANDRA 
(Seward et al.\ 2001, Weis et al.\ 2001), and XMM-Newton results 
(Leutenegger et al.\ 2003) and the notion of $\eta$ Carina being a very 
massive evolved star (but for a different interpretation  see 
e.g.\  Leutenegger et al.\ 2003).
We found indications of very hot gas due to fast shocks and 
detected a previously unseen cloud of soft X-ray emitting gas in front 
of the Homunculus, most probably another cloud in the outer ejecta. 
This cloud underlines the 3d structure of the outer ejecta and that the 
apparent hook shape of the X-ray nebula is only a complex mirage due to 
projection, absorption and true density inhomogeneities.

\begin{acknowledgements}

We thank the referee, Dr. Wolfgang Duschl (Heidelberg), for his helpful
sugesstions  and a fast response. 
Based partly on observations made with the NASA/ESA Hubble Space Telescope,
obtained from the data archive at the Space Telescope Institute. STScI is
operated by the association of Universities for Research in Astronomy,
Inc. under the NASA contract  NAS 5-26555.
This research has made use of NASA's Astrophysics Data System.

\end{acknowledgements}

\end{document}